\documentclass[12pt]{article}
\usepackage{latexsym, amsmath, amsthm, amsfonts, amssymb}
\usepackage{amstext}
\usepackage[all]{xy}
\usepackage{thmtools}
\def\Z{\mathbb{Z}}
\def\R{\mathbb{R}}
\def\T{\mathbb{T}}
\def\act#1#2{#1/\!\!/#2}
\def\AUT{\mathrm{AUT}}

\def\pr{pr}
\def\id{\mathrm{id}}
\def\TBB{\mathbb{TB}^{2}}
\def\TB{\mathbb{TB}^{1}}
\def\TD{\mathbb{TD}}
\def\TFgeo{\mathbb{TD}^{\frac{1}{2}\text{-}geo}}
\def\man{\mathcal{M}an}
\def\cat{\mathcal{C}at}

\def\des#1#2{\mathcal{D}esc_{#1}(#2)}
\def\bun#1{\mathcal{B}un_{#1}}
\def\buncon#1{\mathcal{B}un^{\nabla}_{#1}}
\def\grb#1{\mathcal{B}\mathcal{G}rb_{#1}}
\def\liegrpd{\mathcal{LG}rpd}
\def\hat#1{\widehat{#1}}

\begin{document}
\begin{flushright}
   {\sf ZMP-HH/18-12}\\
   {\sf Hamburger$\;$Beitr\"age$\;$zur$\;$Mathematik$\;$Nr.$\;$740}\\[2mm]
   July 2018
\end{flushright}
\vskip 3.5em

\begin{center}
\begin{tabular}c \Large\bf
Non-abelian gerbes and some applications in string theory
\end{tabular}
\vskip 2.6em

  Christoph Schweigert,$^{\,a}$,
  ~~and~~~ Konrad Waldorf \,$^{\,b}$

\vskip 9mm

  \it$^a$
 Fachbereich Mathematik, \ Universit\"at Hamburg\\
  Bereich Algebra und Zahlentheorie\\
  Bundesstra\ss e 55, \ D\,--\,20\,146\, Hamburg \\[7pt]
  \it$^b$
  Institut f\"ur Mathematik und Informatik, Universit\"at Greifswald \\
  Walther-Rathenau-Stra\ss e 47, \ D\,--\,17\,487\, Greifwald
\end{center}

\vskip 1.3em

\noindent{\sc Abstract}\\[3pt]
We review a systematic construction of the 2-stack of bundle gerbes
via descent, and extend it to non-abelian gerbes. We review the
role of non-abelian gerbes in orientifold sigma models, for the
anomaly cancellation in supersymmetric sigma models, and in a
geometric description of so-called non-geometric T-duals.

\newpage

\section{Introduction}

Higher structures are an important recent trend in mathematics. 
They arise in many fields, most notably in representation theory
and in geometry.
In geometric applications, one considers not only classical
geometric objects, e.g.\ manifolds and fibre bundles on them, but
also objects of a higher categorical nature.
In this contribution, we explain why higher structures naturally 
appear in string theory and, more generally,
in sigma models.

Various categories of manifolds (with  additional structure) appear in string theory:
bosonic sigma models can be defined on smooth manifolds
with a metric; for fermions,
a spin structure (and even a string structure) has to be chosen.
Symplectic manifolds appear in the discussion
of A-models, and complex manifolds for B-models of topologically twisted
string backgrounds.  Already in the very early days of string
theory, it was clear that one should go beyond manifolds to get more interesting
classes of models: in the orbifold construction, one considers manifolds
with a group action that is not necessarily free. 
In modern language, an orbifold is a  proper \'etale Lie groupoid and
thus an object of a bicategory. Thus, in a certain sense, orbifolds
can be seen as a first instance of a higher structure in string theory.

Another source of higher structures in string theory are $p$-form gauge
fields. Bosonic string theory has
a 2-form gauge field, the Kalb-Ramond field. It comes with gauge 
transformations parameterized by 1-form gauge fields, and there are gauge transformations
of gauge transformations parameterized by $U(1)$-valued functions. 
Later, Ramond-Ramond fields in superstring theory have been a rich
source of gauge fields associated to forms of higher degree. The modern
framework to describe such gauge fields are gerbes and
their higher categorical generalizations. 
They have played an
important role in this project of the SFB 676. 

The appropriate mathematical framework for the description
of ordinary 1-form gauge fields are principal $U(1)$-bundles with connection.
Given such a connection, one obtains parallel
transport, and a holonomy map that assigns to a closed curve
an endomorphism of the fibre, which can be
identified with a group element. This holonomy enters in the
action functional of charged particles, and plays an important role e.g.\ in the discussion of
the Aharonov-Bohm effect.

The framework for 2-form gauge fields are $U(1)$-gerbes with connection. Various concrete realizations are available: Dixmier-Douady sheaves of groupoids, \v Cech-Deligne cocycles, bundle gerbes, principal 2-bundles etc. In this contribution we will focus on bundle gerbes. A connection on a  $U(1)$-bundle gerbe leads
to surface holonomy, i.e.\ it associates to a closed oriented
surface an element in $U(1)$. This provides a rigorous geometric
description of Wess-Zumino terms in general target space topology. Similarly, connections on higher bundle gerbes lead
to a notion of holonomy for higher-dimensional
manifolds. The Chern-Simons term in three-dimensional field theories,
for example, can be interpreted as the holonomy of a 2-gerbe.
A review concentrating on surface holonomy of gerbes and their applications in string theory has been written by the authors, together with J\"urgen Fuchs and Thomas Nikolaus, based on earlier results obtained in this project in the SFB 676 \cite{fuchs8}.

In this contribution, we focus on another, equally important aspect of higher geometry, 
namely its treatment in the framework of higher stacks. In a certain sense, this exhibits 
geometric objects as local objects. Some aspects of the theory of higher stacks 
(mostly: 2-stacks), as well as a number of applications to string theory  have been 
developed within this project of the SFB 676.

We start  in Section \ref{plus} by a gentle introduction to stacks and 2-stacks, and provide a new and conceptually clear definition of the bicategory of abelian bundle gerbes. In Section \ref{nonabelian}, we generalize this definition to non-abelian bundle gerbes, clarifying many open issues in an elegant way. In Section \ref{Liegroupoids}  we show that our treatment in the framework of 2-stacks has the additional advantage that it automatically extends the whole theory from manifolds to Lie groupoids. In particular, equivariant (non-abelian) gerbes are a canonical output of this approach. In Section \ref{jandl} we study Jandl gerbes, the gauge fields in orientifold sigma models, as non-abelian gerbes over certain action groupoids. 
This allows a systematic approach to Jandl gerbes.
In Section \ref{string} we report some recent results about string structures on manifolds, and their applications to supersymmetric sigma models, with an emphasis on a description of string structures by non-abelian gerbes for the string group. 
Finally, in Section \ref{tduality}, we describe another application of non-abelian gerbes in the context of T-duality.  The 2-stack-theoretical properties of non-abelian gerbes are essential in this application: we glue locally defined T-duality correspondences to obtain a globally defined new object, a version of a so-called T-fold.

\section{A new perspective to bundle gerbes}

\label{plus}

A hallmark of any geometric theory is  the possibility
to obtain global objects from locally defined objects by a gluing procedure.
In this way, globally defined geometric objects keep aspects of locality.
The essential information  for gluing is the categorical structure of the local model. For instance, principal $G$-bundles 
(with $G$ a finite-dimensional Lie group)
can be glued from trivial bundles along $G$-valued transition functions. In this example, the local model is a category with a single object (the trivial $G$-bundle), whose morphisms are smooth $G$-valued maps (the automorphisms of the trivial bundle). All information is contained in this local model; 
in this case, even, only in its morphisms.
The local   model should be contravariant in the base manifold, so that one can restrict to smaller subsets. In more technical terms, it should form a \emph{presheaf of categories} over the category of smooth manifolds: a (weak) functor
\begin{equation*}
\mathfrak{X}: \man^{op} \to \cat\text{.} 
\end{equation*} 
It includes the assignment of a category $\mathfrak{X}(M)$ to each smooth manifold $M$, and of a functor $f^{*}: \mathfrak{X}(N) \to \mathfrak{X}(M)$ to each smooth map $f: M \to N$, compatible with the composition of smooth
maps in a certain way. In the example of principal $G$-bundles, we have  $\mathfrak{X}(M) = BC^{\infty}(M,G)$, using the notation $BG$ for the category with a single object whose automorphism group is $G$. The functor $f^{*}: BC^{\infty}(N,G) \to BC^{\infty}(M,G)$ is trivial on the level of objects, and $g \mapsto f \circ g$ on the level of morphisms.

Usually (and so in our example) local models do not contain the global objects. In other words, the gluing of local objects will not produce another 
local object.  
In more technical terms, the presheaves $\mathfrak{X}$ are usually not \emph{sheaves of categories}, or \emph{stacks}. The notion of a stack depends, in the first place, on a notion of locality for the underlying category. Essentially,  one has to specify a class of morphisms that correspond to open covers in an abstract sense. 
In case of $\man$, one may consider all maps of the form
\begin{equation}
\label{cover}
\coprod_{i\in I} U_i \to M: (i,x) \mapsto x\text{,}
\end{equation}
where $(U_i)_{i\in I}$ is an open cover of $M$. We shall be more specific about what we want to glue. Given an open cover $U=(U_i)_{i\in I}$ of $M$, we consider a collection $(X_i)_{i\in I}$ of local objects $X_i \in \mathfrak{X}(U_i)$, together with a collection of isomorphisms 
$g_{ij}:X_i|_{U_i \cap U_j} \to X_j|_{U_i \cap U_j}$ in $\mathfrak{X}(U_i \cap U_j)$, for all two-fold overlaps,
which are compatible in the sense that a cocycle condition
\begin{equation}
\label{cocycle}
g_{jk} \circ g_{ij} = g_{ik}    
\end{equation}
is satisfied in $\mathfrak{X}(U_i \cap U_j \cap U_k)$, for all three-fold overlaps.
A pair $((X_i),(g_{ij}))$ is called a \emph{descent object} for the presheaf $\mathfrak{X}$ with respect to the cover $U$. There is a natural notion of morphisms between descent 
objects, so that a category $\des{\mathfrak{X}}{U}$ is formed.

If the gluing of a descent object  $((X_i),(g_{ij}))$  within $\mathfrak{X}$ could be performed, then it is expected to result in a global object $X\in \mathfrak{X}(M)$ that locally restricts to the given local objects $X_i$ in a way compatible with the gluing isomorphisms $g_{ij}$. More precisely, we associate to $X$ the descent object $X_U:=(X|_{U_i},\id)$ and require that $(X|_{U_i},\id) \cong ((X_i),(g_{ij}))$ in $\des{\mathfrak{X}}{U}$. In other words, we consider the functor
\begin{equation}
\label{res}
\mathfrak{X}(M) \to \des{\mathfrak{X}}{U} : X \mapsto X_U
\end{equation}
and demand it to be essentially surjective. For the definition of a stack we require even a bit more, in order to be able to glue not only objects but also morphisms: a presheaf $\mathfrak{X}$ of categories is called a \emph{stack}, if the functor (\ref{res}) is an equivalence of categories for every open cover $U$.

As mentioned before, typical local models do not form stacks but only so-called \emph{prestacks}, meaning that the functors (\ref{res}) are not essentially surjective, but still fully faithful. However, there is a procedure, called \emph{plus construction}, to turn a prestack $\mathfrak{X}$ into a stack  $\mathfrak{X}^{+}$. The idea is very simple: an object in $\mathfrak{X}^{+}(M)$ is a pair of an open cover $U$ and an object in $\des{\mathfrak{X}}{U}$. The morphisms are defined over common refinements of open covers, and we shall omit the details here. 
One can show that the plus construction is idempotent for prestacks, i.e. $\mathfrak{X}^{++}\cong \mathfrak{X}^{+}$ \cite{Nikolaus:2010vq}. 
In our example of the local model $\mathfrak{X}=BC^{\infty}(-,G)$ for principal $G$-bundles, one can then check that $\mathfrak{X}^{+}(M)$  is canonically equivalent to the usual category $\bun G(M)$ of principal $G$-bundles over $M$, via the clutching construction.

One can  admit more general morphisms in our notion of locality, for example one can consider all surjective submersions $\pi:Y \to M$ instead of just the one of the form (\ref{cover}). This has for example advantages in the construction of bundle gerbes
on compact Lie groups.
The $k$-fold intersections are then replaced by the $k$-fold fibre product $Y^{[k]} = Y \times_M ... \times_M Y$, and the definition of the descent category $\des{\mathfrak{X}}\pi$ is fully analogous. In fact, it turns out that the resulting stackification $\mathfrak{X}^{+}$ will be the same as before \cite{Nikolaus:2010vq}.

The main advantage of this approach to geometrical objects over manifolds is that it is very general: its only input is the local model $\mathfrak{X}$. For example,  we let $\mathfrak{X}$ be the following prestack: the objects of $\mathfrak{X}(M)$ are all $\mathfrak{g}$-valued 1-forms on $M$, and the morphisms are all gauge transformations. Then, $\mathfrak{X}^{+}$ is the stack $\buncon G$ of principal $G$-bundles with connections. On the other hand, this approach does often not bring much new insight into the geometric objects themselves, since nice geometric models are well-known, as in the cases of principal $G$-bundles with and without connections.

Analogous considerations persist in higher-categorical settings with more interesting results \cite{Nikolaus:2010vq}. For instance, in one categorical degree higher we consider presheaves $\mathfrak{X}$ of bicategories as our local models. Now, the definition of the descent category has to be changed in order to incorporate 2-morphisms. A descent object in $\des{\mathfrak{X}}{U}$ is a collection of objects $X_i$ in $\mathfrak{X}(U_i)$ and of morphisms $g_{ij}$ in $\mathfrak{X}(U_i\cap U_j)$ 
as before, but instead of the cocycle condition (\ref{cocycle}) it includes additionally a collection of 2-isomorphisms
\begin{equation*}
\mu_{ijk}: g_{jk} \circ g_{ij} \Rightarrow g_{ik} 
\end{equation*} 
in $\mathfrak{X}(U_i\cap U_j \cap U_k )$ satisfying a new cocycle condition in $\mathfrak{X}(U_i\cap U_j\cap U_k \cap U_l)$. Again, there is a 2-functor
\begin{equation*}
\mathfrak{X}(M) \to \des{\mathfrak{X}}U
\end{equation*}
and we call $\mathfrak{X}$ a \emph{pre-2-stack} if  -- for all open covers $U$ -- it induces an equivalence on  Hom-categories, and a \emph{2-stack} if it is an equivalence of bicategories. A generalization of the plus construction produces a 2-stack $\mathfrak{X}^{+}$ out of any pre-2-stack $\mathfrak{X}$ \cite{Nikolaus:2010vq}. 
A concrete example, and the main motivation for this section, is a local model where the morphisms  are principal $U(1)$-bundles. We consider the pre-2-stack $\mathfrak{X}=B\bun{U(1)}$ : the bicategory $\mathfrak{X}(M)$ has a single object, its automorphisms are principal $U(1)$-bundles over $M$, and the 2-morphisms are all bundle morphisms. The composition is the tensor product of principal $U(1)$-bundles. 
Again, all information is in the morphisms, this time in the morphism categories.
The reader is encouraged to perform a brief check, that the plus construction (with respect to surjective submersions) produces exactly the bicategory of $U(1)$-bundle gerbes,
\begin{equation*}
\grb{U(1)}(M) = (B\bun{U(1)})^{+}(M)
\end{equation*}
as described by Murray and Stevenson \cite{murray,murray2,stevenson1}.  Let us summarize three advantages of this result:
\begin{enumerate}

\item 
It derives the definition of a bundle gerbe from first principles.

\item
It automatically produces the quite complicated bicategorical structure of bundle gerbes, whose development by hand took many years.

\item
By construction, $\grb{U(1)}$ is a 2-stack, i.e. bundle gerbes can be glued. Variants of this result have been proved by hand by Stevenson \cite{stevenson1} and Meinrenken \cite{meinrenken1}.

\end{enumerate}

It is straightforward to find interesting variations. For example, one can take any abelian Lie group $A$ instead of $U(1)$, since then $\bun A$ is still a monoidal category. One can take 
hermitian line bundles instead of principal $U(1)$-bundles, resulting in a line bundle version of bundle gerbes. One can even consider hermitian vector bundles of \emph{higher} (but finite) rank. The reader is again encouraged to check that this does not change the definition of a single bundle gerbe, but it produces a  bicategory with more 1-morphisms, in particular, non-invertible ones. Finally, one can add connections to the picture, and thus consider a pre-2-stack $\mathfrak{X}$ whose bicategory $\mathfrak{X}(M)$ is the following:
\begin{itemize}

\item 
Its objects are 2-forms $B \in \Omega^2(M)$. 

\item
The Hom-category between objects $B_1$ and $B_2$ is the full subcategory of hermitian vector bundles $E$ over $M$ with unitary connections $\nabla$ of curvature
\begin{equation}
\label{eq:curvcond}
\frac{1}{\mathrm{rk}(E)}\,\mathrm{tr}(\mathrm{curv}(\nabla))=B_2-B_1\text{.}
\end{equation}

\end{itemize}
The plus construction results precisely in the bicategory of bundle gerbes with connection described in \cite{Waldorf:2007}. 
The proof of the fact that bundle gerbes with connection form a 2-stack should be seen as
one important mathematical result of this project of the SFB676.

We remark  that bundle gerbes with connection give rise to a notion of surface holonomy. Surface holonomy is the basis of many applications of bundle gerbes in string theory: connections on bundle gerbes are the Kalb-Ramond gauge fields for strings, and surface holonomy provides the coupling term in the string action. The extension from line bundles to vector bundles allows to discuss twisted Chan-Paton gauge fields on D-branes within this framework \cite{Waldorf:2007}. For a more detailed discussion of surface holonomy we refer to our review \cite{fuchs8}.

\section{Non-abelian gerbes}

\label{nonabelian}

In this section we demonstrate the full power of the plus construction in the case of so-called non-abelian gerbes. The terminology is not totally accurate, as the abelian group $U(1)$ of the previous section is generalized to a \emph{Lie 2-group} (instead of a non-abelian group). Non-abelian gerbes have found several possible applications in string theory:
\begin{itemize}

\item
Connections on non-abelian gerbes are the gauge fields in higher gauge theory; see \cite{Baez2011} for an overview. In M-theory, they arise  in the Lagrangian description of M5-branes, see \cite{Saemann2014,Fiorenza2014,Jurco2015}.

\item
Non-abelian gerbes have been used to describe 4-dimensional topological quantum field theories that control the long-distance behaviour of surface operators in gapped phases of 4-dimensional gauge theories \cite{Gukovb}.

\item 
Graded $U(1)$-gerbes can be seen as non-abelian gerbes; they appear in orientifold sigma models (see Section \ref{jandl} and the references therein), and as geometrical models for twistings of K-theory \cite{Freed2016a}.

\item
Non-abelian gerbes for the string 2-group are relevant for the  anomaly cancellation in supersymmetric sigma models, see Section \ref{string} and the references therein.

\item
There is a reformulation of topological T-duality in terms of non-abelian gerbes provides a geometric description of so-called non-geometric T-duals, see Section \ref{tduality} and \cite{Nikolause}.

\end{itemize}

The definition of a non-abelian bundle gerbe has been extrapolated manually from the abelian case \cite{aschieri,breen1}, but can be derived systematically via the plus construction \cite{Nikolaus}. Before starting to describe the second approach, we shall explain the notion of a \emph{Lie 2-group}, which plays the role of the ,,structure group`` of a non-abelian gerbe.

A Lie 2-group is a Lie groupoid with a certain kind of monoidal structure. A Lie 2-group is called \emph{strict}, if the monoidal structure is strictly associative, i.e. its associator is trivial. We will concentrate on the strict case. Strict Lie 2-groups can equivalently be described by \emph{crossed modules} of Lie groups. A crossed module is a Lie group homomorphism $t:H \to G$ together with an action $\alpha$ of $G$ on $H$ by group homomorphisms, such that
\begin{equation*}
\alpha(t(h),x)=hxh^{-1}
\quad\text{ and }\quad
t(\alpha(g,h))=gt(h)g^{-1}\text{.}
\end{equation*} 
The Lie groupoid that corresponds to such a crossed module has objects $\Gamma_0 := G$ and morphisms $\Gamma_1 := H \times G$, with source map $(h,g) \mapsto g$ and target map $(h,g)\mapsto t(h)g$. The composition is given by the group structure of $H$, and the monoidal structure is given by the group structures of $G$ and  of the semi-direct product $H \ltimes G$ formed using the action $\alpha$ of $G$ on $H$. The standard examples of Lie 2-groups are the following; more examples will be mentioned in Sections \ref{string} and \ref{tduality}.
\begin{itemize}

\item 
For an abelian Lie group $A$, there is a Lie 2-group $BA$ with a single object and the group $A$ as its automorphisms; this construction (and notation) is analogous to Section \ref{plus}. Composition and monoidal structure are both given by the group structure of $A$;  since the monoidal structure is a functor, the classical Eckmann-Hilton argument requires $A$ to be abelian. The corresponding crossed module is $A \to \{e\}$.

\item
For a general Lie group $G$, there is a Lie 2-group $G_{dis}$ with objects given by the elements of $G$ 
and only identity morphisms. The corresponding crossed module is $\id:G \to G$, together with the conjugation of $G$ on itself. 

\item
If the automorphism group $\mathrm{Aut}(H)$ of a Lie group $H$ is again a Lie group (for example, it is discrete when $H$ is compact and simple), then there is a Lie 2-group $\mathrm{AUT}(H)$ called the automorphism 2-group of $H$. Its crossed module is the assignment $i: H \to \mathrm{Aut}(H)$ of  inner automorphisms, together with the natural action of $\mathrm{Aut}(H)$ on $H$. 

\end{itemize}

We explain two more facts about Lie 2-groups that will be relevant later. Lie 2-groups have two interesting invariants, $\pi_0\Gamma$ (the group of isomorphism classes of objects) and $\pi_1\Gamma$ (the abelian group of automorphisms of the monoidal unit).  There is  an action of $\pi_0\Gamma$ on $\pi_1\Gamma$ obtained by conjugation with identity morphisms, and the Lie 2-group is called \emph{central} if this action is trivial.  In terms of crossed modules, $\pi_0\Gamma = G/t(H)$ and $\pi_1\Gamma=\mathrm{ker}(t)\subset H$, and the action is induced from $\alpha$. For example, $\pi_0\mathrm{AUT}(H)=\mathrm{Out}(H)$ and $\pi_1\mathrm{AUT}(H)=Z(H)$, and $\mathrm{AUT}(H)$ is central if and only if every outer automorphism fixes the center.

A Lie 2-group is called \emph{smoothly separable}, if $\pi_0\Gamma$ is a Lie group such that $\Gamma_0 \to \pi_0\Gamma$ is a submersion. Every smoothly separable Lie 2-group gives rise to an extension
\begin{equation}
\label{extension}
B\pi_1\Gamma \to \Gamma \to (\pi_0\Gamma)_{dis}
\end{equation}
of Lie 2-groups  in the sense of Schommer-Pries \cite{pries2}. If $\Gamma$ is central then this extension is  central.

In order to specify the input data for the plus construction, we have to specify a local model for non-abelian gerbes. The idea is analogous to the abelian case: we want to define a monoidal category of bundles related to the given Lie 2-group $\Gamma$. 
For any Lie groupoid $\Gamma$ one can consider principal $\Gamma$-bundles, and we refer to \cite{Nikolaus} for a review. Principal $\Gamma$-bundles are ordinary fibre bundles $\pi:P \to M$, whose total spaces are equipped with an ,,anchor map`` $\phi:P \to G$ and an action of $\Gamma$ -- this means that a morphism $\gamma:a \to b$ acts on  points $p\in P$ with anchor $\phi(p)=a$, resulting in a new point $p\circ \gamma$ with anchor $b$, in the same fibre over $M$. The action is supposed to satisfy the usual conditions for principal bundles. 
The monoidal structure on $\Gamma$ is used for the definition of the tensor product of principal $\Gamma$-bundles. 
 
In terms of a crossed module $t:H \to G$, a principal $\Gamma$-bundle is an ordinary principal $H$-bundle $P$ equipped with a smooth, anti-equivariant map $\phi:P\to G$, i.e. $\phi(p\cdot h)=t(h)^{-1}\cdot\phi(p)$. The tensor product $P \otimes Q$ is the fibre product of the underlying bundles, $P \times_M Q$, modulo an equivalence relation $(ph,q)\sim (p,q\alpha(\phi_P(p)^{-1},h))$, with the $H$-action defined by $[p,q]h=[ph,q]$ and the anti-equivariant map defined by $[p,q]\mapsto \phi_P(p)\phi_Q(q)$.

The reader may easily verify that principal $\Gamma$-bundles for the three Lie 2-groups of the previous list are the following:
\begin{itemize}

\item 
Principal $BA$-bundles are the same as ordinary principal $A$-bundles, with the ordinary tensor product. 

\item
Principal $G_{dis}$-bundles are the same as smooth maps $M \to G$, with the pointwise group structure.

\item
Principal $\Gamma=\mathrm{AUT}(H)$-bundles are ordinary 
principal $H$-bundles $P$ together with an $H$-anti-equivariant map $\phi: P \to \mathrm{Aut}(H)$. Due to the anti-equivariance, one can define an additional \emph{left} $H$-action $h p := p\alpha(\phi(p)^{-1},h)$, turning $P$ into a $H$-bibundle. The tensor product is the tensor product of $H$-bibundles. 
 The theory of non-abelian gerbes started with considering  bibundles and corresponding $\mathrm{AUT}(H)$-bundle gerbes \cite{aschieri,breen1}. 

\end{itemize}

So far we have explained the monoidal category $\bun\Gamma (M)$ of principal $\Gamma$-bundles over $M$. The local model for $\Gamma$-bundle gerbes is now the pre-2-stack $BBun_{\Gamma}$.  
The plus construction yields then the 2-stack of $\Gamma$-bundle gerbes,
\begin{equation*}
\grb{\Gamma} := (B\bun\Gamma)^{+}\text{.}
\end{equation*}

Unpacking the details of the plus construction, a $\Gamma$-bundle gerbe consists of a surjective submersion $\pi:Y \to M$, a principal $\Gamma$-bundle $P$ over $Y^{[2]}$, and a bundle isomorphism
\begin{equation*}
\mu: pr_{23}^{*}P \otimes pr_{12}^{*}P \to pr_{13}^{*}P
\end{equation*}
over $Y^{[3]}$ that satisfies a cocycle condition over $Y^{[4]}$.

We would like to emphasize that although non-abelian bundle gerbes have been defined earlier, e.g. in \cite{aschieri}, neither the full bicategorical structure has been specified, nor any gluing properties have been discussed. Both important aspects are established automatically by the plus construction. A further important aspect that comes for free from the plus construction is the functoriality in the Lie 2-group $\Gamma$. That is, if $F: \Gamma \to \Omega$ is a Lie 2-group homomorphism, then there is an associated ,,change of structure 2-group`` 2-functor
\begin{equation*}
F_{*}: \grb\Gamma(M) \to \grb\Omega(M)\text{.}
\end{equation*}
In particular, as a consequence of the extension (\ref{extension}),
if $\Gamma$ is smoothly separable with $A:=\pi_1\Gamma$ and $G:=\pi_1\Gamma$, then there is a sequence
\begin{equation}
\label{lifting}
\grb{A}(M) \to \grb{\Gamma}(M) \to (\bun G(M))_{dis}
\end{equation}
of 2-functors. Here we have employed an identification between (non-abelian) $BA$-bundle gerbes and (abelian) $A$-bundle gerbes, and another identification  $\grb{G_{dis}}=(\bun G)_{dis}$ between $G_{dis}$-bundle gerbes and ordinary principal $G$-bundles (regarded as a bicategory with only identity 2-morphisms). Loosely speaking, the sequence (\ref{lifting}) exhibits non-abelian gerbes as an extension of ordinary principal bundles by abelian gerbes.  In particular, every non-abelian $\Gamma$-bundle gerbe $\mathcal{G}$ comes with an underlying ordinary principal $\pi_0\Gamma$-bundle, which we denote by $\pi_0(\mathcal{G})$. This bundle plays an important role in the applications, as we will see in Sections \ref{jandl}, \ref{string} and \ref{tduality}. In work with Thomas Nikolaus we have studied several lifting and reduction statements for non-abelian gerbes that are related to the sequences (\ref{extension}) and (\ref{lifting}), see \cite{Nikolausa}.

We remark that connections on non-abelian bundle gerbes can be defined in the very same way via the plus construction. The generalization of surface holonomy to the non-abelian case is more difficult. In joint work with Urs Schreiber we have given a general and axiomatic framework for parallel transport and holonomy of non-abelian gerbes \cite{schreiber3,schreiber5,schreiber6,schreiber2}.

We also point out that there is a formalism of \emph{principal 2-bundles}, initiated by Bartels \cite{bartels} and Wockel \cite{Waldorf2016}. This formalism is equivalent to non-abelian bundle gerbes \cite{Nikolaus} but more suitable  for connections and parallel transport, see \cite{Waldorf2016,Waldorf2017}.

\section{Gerbes over Lie groupoids}

\label{Liegroupoids}

Applications (in particular applications to sigma-models and string theory)
frequently require not only bundle gerbes, but \emph{equivariant} bundle gerbes. It is fruitful to approach equivariant geometry from a more general point of view: geometry over \emph{Lie groupoids}. The guiding examples of Lie groupoids are the following two:
\begin{itemize}

\item
If a Lie group $G$ acts  on a smooth manifold $M$ in terms of a smooth map $\rho:G \times M \to M$, one can form the \emph{action groupoid} $\act MG$ with objects $(\act MG)_0:= M$ and morphisms $(\act MG)_1 := G \times M$. Source and target maps are given by $s(g,m):= m$ and $t(g,m):=\rho(g,m)$, and the composition is $(g_2,g_1m)\circ (g_1,m) := (g_2g_1,m)$. Geometry over an action groupoid $\act MG$ will be the same as $G$-equivariant geometry over $M$.

\item 
Any open cover $U=(U_i)_{i\in I}$ defines the so-called \emph{\v Cech groupoid} $\check C(U)$. Its objects and morphisms are given by, respectively,
\begin{equation*}
\check C(U)_0 := \coprod_{i\in I} U_i
\quad\text{ and }\quad
\Check C(U)_1 := \coprod_{i,j\in I} U_i \cap U_i\text{.}
\end{equation*}
Source and target maps are given by $s(i,j,x):=(i,x)$ and $t(i,j,x):=(j,x)$, and the composition is $(j,k,x)\circ (i,j,x) := (i,k,x)$. A similar Lie groupoid $\check C(\pi)$ can be constructed 
using fibre products for any surjective submersion $\pi:Y \to M$.

\end{itemize}

We remark that a further class of interesting and rich examples of Lie groupoids are 
orbifolds, see 
\cite{lerman1,Moerdijk1997,Moermrc}.
We start again by describing presheaves of categories; now over Lie groupoids.
A \emph{presheaf of categories over Lie groupoids} is a weak functor
\begin{equation*}
\mathfrak{X}: \liegrpd^{op} \to \cat\text{,}
\end{equation*}
i.e., it associates to each Lie groupoid $\Omega$ a category $\mathfrak{X}(\Omega)$, and 
each smooth functor $F: \Omega \to \Omega'$ of Lie groupoids
a functor $F^{*}:\mathfrak{X}(\Omega') \to \mathfrak{X}(\Omega)$ in  a way compatible with the composition of functors. There is a canonical way to extend any presheaf $\mathfrak{X}$  over smooth manifolds to a presheaf $\mathfrak{X}'$ over Lie groupoids \cite{Nikolaus:2010vq}. Indeed, if $\Omega$ is a Lie groupoid with a manifold $\Omega_0$ of objects and a manifold $\Omega_1$ of morphisms, then an object of $\mathfrak{X}'(\Omega)$ is a pair $(X,f)$ consisting of an object $X$ 
in $\mathfrak{X}(\Omega_0)$ and an isomorphism $f: s^{*}X \to t^{*}X$ in $\mathfrak{X}(\Omega_1)$, such that $\pr_2^{*}f \circ \pr_1^{*}f=c^{*}f$ as morphisms in $\mathfrak{X}(\Omega_1 \, _{s}\!\times_t \Omega_1)$, where $c$ denotes the composition. A morphism of $\mathfrak{X}'(\Omega)$ between $(X,f)$ and $(X',f')$ is a morphism $g: X \to X'$ in $\mathfrak{X}(\Omega_0)$ such that $f'\circ s^{*}g = t^{*}g \circ f$ in $\mathfrak{X}(\Omega_1)$.

It is instructive to evaluate this procedure for the two examples of Lie groupoids described above. An object in $\mathfrak{X}'(\act MG)$ is an object $X$ over $M$ together with an isomorphism $f: \pr_M^{*}X \to \rho^{*}X$ over $G \times M$ that satisfies above condition over $G \times G \times M$. In other words, this is a family 
$\{f_g\}_{g\in G}$ of isomorphisms $f_g: X \to g^{*}X$ that satisfy $g_1^{*}f_{g_2}\circ f_{g_1}=f_{g_2g_1}$ and depend smoothly on $G$ (in the sense that a morphism $f$ over $G \times M$ is formed). If, for example, $\mathfrak{X}$ is the stack of vector bundles, then an object in $\mathfrak{X}'(\act MG)$ is precisely a $G$-equivariant vector bundle over $M$.

In the other example of \v Cech groupoids,
the reader may easily verify that $\mathfrak{X}'(\check C(\pi))=\des {\mathfrak{X}}\pi$, for any surjective submersion $\pi:Y \to M$. This observation is useful in the following situation. Suppose the quotient of a   $G$-action on $M$ exists, in the sense that $M/G$ is a smooth manifold and $p:M \to M/G$ is a principal $G$-bundle. Then, there is a canonical isomorphism of Lie groupoids $\check C(p) \cong \act MG$. Thus, if $\mathfrak{X}$ is any stack over smooth manifolds, we have an equivalence 
\begin{equation}
\label{equivdesc}
\mathfrak{X}(M/G) \cong \des {\mathfrak{X}}p = \mathfrak{X}'(\check C(p)) \cong \mathfrak{X}'(\act MG)\text{.}
\end{equation}
In other words, if the quotient exists, geometry over $M/G$ is the same as the induced geometry over the action groupoid. The latter, however, makes sense even if the quotient does not exist.

The passage $\mathfrak{X} \mapsto \mathfrak{X}'$ from presheaves  over smooth manifolds to presheaves over Lie groupoids has the feature that it is functorial and preserves stacks \cite{Nikolaus:2010vq}. It is a very convenient tool: once we work with presheaves of categories, there is no need to introduce definitions over Lie groupoids. These will follow automatically via $\mathfrak{X} \mapsto \mathfrak{X}'$  from definitions over just smooth manifolds.

The above discussion generalizes in a straightforward way from presheaves of \emph{categories} to presheaves of \emph{bicategories} \cite{Nikolaus:2010vq}, and hence applies to abelian and non-abelian bundle gerbes.
Hence, we automatically obtain a definition of equivariant non-abelian bundle gerbes, together with the equivalence (\ref{equivdesc}) saying that equivariant bundle gerbes descent to quotients (if these exist).

Equivariant bundle gerbes are frequently used in 
two-dimensional
Wess-Zumino-Witten models, whose Wess-Zumino term is the holonomy of a $U(1)$-bundle gerbe on a   Lie group. In order
to get the right invariances for the theory, this gerbe has to be
equivariant with respect to the adjoint action of the Lie group on itself.
Concrete constructions of  equivariant $U(1)$-bundle gerbes over compact simple Lie groups use descent in the 2-stack $\grb{U(1)}'$  twice: In the first step, a  $G$-equivariant bundle gerbe (,,basic gerbe``) over a simply-connected  Lie group $G$ is obtained by gluing  $G$-equivariant bundle gerbes $\mathcal{G}_{i}$, which are locally defined over thickened conjugacy classes $U_i \subset G$. In other words, these are bundle gerbes over  action  groupoids $\act {U_i}G$. More precisely, the locally defined bundle gerbes $\mathcal{G}_i$ are so-called lifting gerbes for certain central extensions of stabilizer subgroups. The gluing uses descent along the functor
\begin{equation*}
\coprod_{i} \act {U_i}G \to \act GG\text{.}
\end{equation*} 
This construction is implicit in Meinrenken's construction \cite{meinrenken1} and appears explicitly in \cite{nikolaus1}. 
In the second step, 
descent to a non-simply-connected quotient $\tilde G:=G/Z$, where $Z\subset Z(G)$, is  performed along $\act GG \to \act{\tilde G}{\tilde G}$. The required $Z$-equivariant structures have been provided Lie-theoretically by Gaw\c edzki-Reis \cite{Gawedzki:2003pm} (without the $G$-equivariance) and recently -- including the $G$-equivariance -- by Krepski \cite{Krepski}.

\section{Jandl gerbes are non-abelian}

\label{jandl}

The surface holonomy of a connection on an abelian bundle gerbe depends on an orientation of the surface $\Sigma$. If the surface is not oriented, or not orientable, then it is a priori not well-defined. Thus, in the application of bundle gerbes as Kalb-Ramond gauge fields in string theory, 
unoriented worldsheets 
 (which naturally appear in string theories of type I)
require additional attention.

The idea is that a change of orientation should be accompanied on the target space side with an involution $k:M \to M$, under which the  gauge field should ,,change its sign``. For a 2-form gauge field $B$, we would require $k^{*}B=-B$ so that the sign obtained from a change of orientation is compensated. Jandl gerbes  generalize this transformation behaviour from 2-form gauge fields to bundle gerbes: they are $U(1)$-bundle gerbes equipped with additional structure relating their pullback along $k$ with the dual (for the opposite sign). It is instructive to understand Jandl gerbes as non-abelian gerbes over Lie groupoids.

We are concerned with a smooth manifold $M$ and an involution $k: M \to M$, which we regard as a $\Z_2$-action on $M$. We consider the associated action groupoid, and denote it by $\act M k$. 
The trivial $\Z_2$-bundle $M \times \Z_2$ over $M$ can be equipped with a $\Z_2$-equivariant structure that changes the sign under the involution. Hence, it becomes a $\Z_2$-bundle over $\act Mk$, and we denote it by $\mathrm{Or}(\act Mk)$, the \emph{orientation bundle} of  the Lie groupoid $\act Mk$.

The automorphism 2-group $\mathrm{AUT}(U(1))$ gives rise to a (non-central) extension
\begin{equation*}
BU(1) \to \mathrm{AUT}(U(1)) \to \Z_2\text{,}
\end{equation*}
where the action of  $\pi_0(\mathrm{AUT}(U(1)))=\mathrm{Out}(\Z_2)=\Z_2$ on $\pi_1(\mathrm{AUT}(U(1)))=Z(U(1))=U(1)$ is by inversion. The idea is to couple the $\Z_2$-part of this extension to a change of orientation. With this motivation, a \emph{Jandl gerbe} over $\act Mk$ is an $\mathrm{AUT}(U(1))$-bundle gerbe $\mathcal{G}$ with connection over $\act Mk$ 
together with a bundle isomorphism $\pi_{0}(\mathcal{G}) \cong \mathrm{Or}(\act Mk)$.

The above definition of a Jandl gerbe is more conceptual than the original definition
given in \cite{Schreiber:2005mi}, 
but equivalent, as we will demonstrate now by unwrapping all involved definitions. We work first over $M$ and ignore the $\Z_2$-equivariance. From the plus construction we recall that the $\mathrm{AUT}(U(1))$-bundle gerbe  $\mathcal{G}$  consists of a surjective submersion $\pi:Y \to M$, a principal $\mathrm{AUT}(U(1))$-bundle $P$ over $Y^{[2]}$, and a bundle isomorphism $\mu$ over $Y^{[3]}$. We consider $\mathrm{AUT}(U(1))$ as the crossed module $0: U(1) \to \Z_2$, so that the $\mathrm{AUT}(U(1))$-bundle $P$ is an ordinary $U(1)$-bundle equipped with an anchor map $\phi:P \to \Z_2$. The anchor descends in fact to a map $\phi:Y^{[2]} \to \Z_2$, due to its anti-equivariance. The bundle isomorphism $\mu$ is an ordinary $U(1)$-bundle isomorphism with the property that it respects the anchor maps; this implies the cocycle condition for the map $\phi$ on $Y^{[3]}$. Hence, the pair $(\pi,\phi)$ is descent data for a principal $\Z_2$-bundle over $M$, namely the bundle $\pi_0(\mathcal{G})$. 

The isomorphism between $\pi_0(\mathcal{G})$ and $\mathrm{Or}(\act Mk)$ that is part of the definition of a Jandl gerbe determines a trivialization of $\pi_0(\mathcal{G})$, since $\mathrm{Or}(\act Mk)$ is the trivial bundle over $M$. In terms of descent data, the trivialization is a map $\psi: Y \to \Z_2$ such that  $\phi(y_1,y_2)\psi(y_2)=\psi(y_1)$.  Let $P_{\psi}$ be the trivial $U(1)$-bundle over $Y$, which becomes an $\mathrm{AUT}(U(1))$-bundle equipping it with the anchor map $\psi$. Now we pass to the new $\AUT(U(1))$-bundle
\begin{equation*}
P_{\mathrm{red}} := \pr_2^{*}P_{\psi} \otimes P \otimes \pr_1^{*}P_{\psi}^{*}
\end{equation*}
over $Y^{[2]}$.
By construction $P_{\mathrm{red}}$ has the trivial anchor and so is an ordinary $U(1)$-bundle. Similarly, one can equip $P_{\mathrm{red}}$ with an isomorphism $\mu_{\mathrm{red}}$ over $Y^{[3]}$, in such a way that $\mathcal{G}_{\mathrm{red}} :=(\pi,P_{\mathrm{red}},\mu_{\mathrm{red}})$ is an ordinary $U(1)$-bundle gerbe. This reduction procedure  works in the same way in the setting with connections. We have studied it in a more general setting in \cite{Nikolausa}. 

Next we take care about the $\Z_2$-equivariant structure, which is -- in the first place -- an isomorphism $\mathcal{A}: s^{*}\mathcal{G} \to t^{*}\mathcal{G}$ of $\mathrm{AUT}(U(1))$-bundle gerbes over the morphism space $\Z_2 \times M$ of the action groupoid $\act Mk$.  It induces an isomorphism
$\mathcal{A}: s^{*}\mathcal{G}_{\mathrm{red}} \to t^{*}\mathcal{G}_{\mathrm{red}}$
over $\Z_2 \times M$, which is an isomorphism in the bicategory of $\mathrm{AUT}(U(1))$-bundle gerbes. That space is the disjoint union of two components, and so $\mathcal{A}$ has two components $\mathcal{A}_{\id}: \mathcal{G}_{\mathrm{red}} \to \mathcal{G}_{\mathrm{red}}$ and $\mathcal{A}_k: \mathcal{G}_{\mathrm{red}} \to k^{*}\mathcal{G}_{\mathrm{red}}$. Employing the equivariance of the isomorphism $\pi_0(\mathcal{G}) \cong \mathrm{Or}(\act Mk)$ one can show that $\mathcal{A}_{\id}$ is actually an isomorphism in the bicategory of $U(1)$-bundle gerbes. For the component $\mathcal{A}_k$ one can show that a sign is involved in such a way that $\mathcal{A}_k$  becomes an isomorphism of $U(1)$-bundle gerbes after its domain bundle gerbe is dualized:
\begin{equation*}
\mathcal{A}_k: \mathcal{G}_{\mathrm{red}}^{*} \to k^{*}\mathcal{G}_{\mathrm{red}}\text{.}
\end{equation*}

Finally, we incorporate the last part of the $\Z_2$-equivariant structure, which is a 2-isomorphism $\varphi: pr_2^{*}\mathcal{A} \circ pr_1^{*}\mathcal{A} \Rightarrow  c^{*}\mathcal{A}$ over the space of pairs of composable morphisms of $\act Mk$. It reduces to 2-isomorphisms
\begin{equation*}
\mathcal{A}_{\id} \circ \mathcal{A}_{\id} \Rightarrow \mathcal{A}_{\id}
\quad\text{ and }\quad 
k^{*}\mathcal{A}_k \circ \mathcal{A}_k^{*} \Rightarrow  \id_{\mathcal{G}_{\mathrm{red}}}\text{.}
\end{equation*}
The first part shows that $\mathcal{A}_{\id}\cong\id$, so that $\mathcal{A}_{\id}$ contains no information. For the second part, the cocycle condition for $\varphi$ implies $k^{*}\varphi_k^{-1} = \varphi_k^{*}$. 

Summarizing, we have seen that a Jandl gerbe $\mathcal{G}$ over $\act Mk$ is the same as:
\begin{enumerate}

\item 
A $U(1)$-bundle gerbe with connection over $M$.

\item
A 1-isomorphism $\mathcal{A}: \mathcal{G}^{*} \to k^{*}\mathcal{G}$.

\item
A 2-isomorphism $\varphi: k^{*}\mathcal{A} \circ \mathcal{A}^{*}  \Rightarrow \id_{\mathcal{G}}$ such that $k^{*}\varphi^{-1} = \varphi^{*}$. 

\end{enumerate}
This is precisely the definition given in \cite{Schreiber:2005mi,Waldorf:2007}. We remark that Jandl gerbes can also be discussed in terms of cocycle data, or (differential) cohomology with coefficients in equivariant sheaves, see \cite{Schreiber:2005mi,gawedzki6,Gawedzki:2008um,waldorf4,Hekmati}.

As intended, Jandl gerbes furnish a notion of surface holonomy  for unoriented surfaces \cite{Schreiber:2005mi}. More precisely, if $\Sigma$ is a possibly unoriented surface, it assigns a well-defined element in $U(1)$ to each  differentiable stack map $\phi: \Sigma \to \act Mk$. This surface holonomy constitutes the contribution of the orientifold Kalb-Ramond field to the sigma model action, see \cite{fuchs8,Schreiber:2005mi}.

In \cite{gawedzki6} we have classified all Jandl gerbes over compact simple Lie groups, and thereby all Wess-Zumino-Witten orientifolds for these groups. The classification problem was solved using equivariant descent along the universal covering group $\tilde G \to G$, employing the fact that $\mathrm{AUT}(U(1))$-bundle gerbes form a 2-stack. Since $G=\tilde G/Z$ for a discrete group $Z \subset Z(\tilde G)$, the essential calculation is the classification of all $Z$-equivariant structures on $\mathrm{AUT}(U(1))$-bundle gerbes over simply-connected Lie groups. In \cite{gawedzki6}, this was reduced to the computation of the group cohomology of the discrete group $Z\ltimes \Z_2$ (where $\Z_2$ acts on $Z$ by inversion) with coefficients in $U(1)$, considered as a $(Z \ltimes \Z_2)$-module in which $\Z_2$ acts by inversion. This calculation has been carried out in \cite{gawedzki6} for all occurring cases of groups $Z$.

In further work \cite{Gawedzki:2008um} we have treated D-branes in orientifolds, using Jandl gerbes. In this picture, D-branes are submanifolds $Q \subset M$ of the target space, equipped with bundle gerbe modules, i.e. 1-morphisms $\mathcal{G}|_Q \to \mathcal{I}$, where $\mathcal{I}$ is the trivial bundle gerbe. If these bundle gerbe modules are equipped with appropriate equivariance with respect to the involution, the coupling term in the open string action functional can again be defined unambiguously. 
A different treatment of D-branes in orientifolds has been studied by Distler, Freed, and Moore in \cite{Distler2011},  in the formalism of twisted K-theory. The relation between the two pictures is that bundle gerbes (with connections)  realize those (differential) twistings of K-theory that correspond to $H^3(M,\Z)$. Non-abelian $\mathrm{AUT}(U(1))$-bundle gerbes, which underly Jandl gerbes,  realize more general twistings that correspond to $H^1(M,\Z_2) \times H^3(M,\Z)$. These correspondences persist if $M$ is replaced by an action groupoid; this way the two approaches can be related.

\section{String structures}

\label{string}

Supersymmetric sigma models include  spinors on the worldsheet with values in the tangent bundle of the target space.   If $M$ is a spin manifold, then the path integral over the spinors can be interpreted as a Berezinian integral, and then rigorously be performed. The result, however, is not a complex number but an element in a complex line. These form a complex line bundle over the space $C^{\infty}(\Sigma,M)$ of all worldsheet embeddings, the \emph{Pfaffian line bundle} of a certain family  of Dirac operators. The supersymmetric sigma model is hence potentially anomalous, and the anomaly is represented by the Pfaffian line bundle. A general treatment of such anomalies was given by Freed and Moore in \cite{freed5}.

The cancellation of this anomaly requires to trivialize the Pfaffian line bundle. Freed showed \cite{Freed1986} that its first Chern class vanishes if the first fractional Pontryagin class of $M$ vanishes,
\begin{equation*}
\textstyle\frac{1}{2}p_1(M)=0 \in H^4(M,\mathbb{Z})\text{.}
\end{equation*}
Such manifolds are called \emph{string manifolds}. The problem is that the vanishing of the Chern class of the Pfaffian line bundle  is not enough to make the fermionic path integral a well-defined section: additionally, a specific trivialization must be provided. For this purpose, it is desirable to interpret the obstruction class $\frac{1}{2}p_1(M)$ in a geometric way. There are (at least) four
different proposals:
\begin{enumerate}

\item
It is the obstruction against lifting the structure group of the free loop space of $M$ from the loop group $LSpin(n)$ to its universal central extension \cite{killingback1,mclaughlin1},
\begin{equation*}
1 \to U(1) \to \widetilde{LSpin(n)} \to LSpin(n)\to 1
\end{equation*}  

\item
It is the obstruction against lifting the structure group of the spin-oriented frame bundle $P_{Spin}M$ of $M$ from $Spin(n)$ to the string group
\begin{equation*}
String(n) \to Spin(n)\text{,}
\end{equation*}
defined as the unique (up to homotopy equivalence) 3-connected covering group \cite{stolz1}. 
\item
It is the ,,level`` of a Chern-Simons  field theory  with target space $M$ \cite{stolz1}.

\item
It is the characteristic class of the Chern-Simons 2-gerbe associated to the spin-oriented frame bundle \cite{carey4,waldorf8}.  

\end{enumerate}
Most interesting for us is (a variant of) version 2. Although the (a priori topological) group $String(n)$ can be realized as a Fr\'echet Lie group \cite{Nikolausb}, it turns out to be more natural to consider it as a Lie 2-group. Concrete models for the String 2-group
have been constructed as a Fr\' echet Lie 2-group  \cite{baez9} and as a diffeological 2-group  \cite{Waldorf}. Both are central and smoothly separable in the sense explained in Section \ref{nonabelian}, and  give rise to a central extension
\begin{equation*}
BU(1) \to String(n) \to Spin(n)_{dis}
\end{equation*}
of (Fr\'echet/diffeological) Lie 2-groups. A \emph{string structure} is a lift of the structure group of $M$ along this central extension. In other words, a string structure is a $String(n)$-bundle gerbe $\mathcal{G}$ over $M$ such that $\pi_0(\mathcal{G})\cong P_{Spin}M$.

By a result of Schommer-Pries \cite{pries2}, central extensions of  $G_{dis}$ by $BU(1)$ are classified by $\mathrm{H}^4(BG,\Z)$, and the string 2-group 
corresponds to  $\frac{1}{2}p_1 \in \mathrm{H}^4(BSpin(n),\Z)$. 
This shows that string structures exist if and only if $\frac{1}{2}p_1(M)=0$, as desired.

Above notion of a string structure is equivalent \cite{Nikolausa} to another definition using version 4. In that version a string structure is a trivialization of the Chern-Simons 2-gerbe \cite{waldorf8}. This has the additional advantage that it is totally finite-dimensional, and that one can  define string connections, together forming a \emph{geometric string structure}. 
The geometric string structures of \cite{waldorf8} are motivated by version 3 of Stolz and Teichner, and closely related to spin structures on the free loop space, which implement version 1 \cite{waldorfb}.

The relation between  string structures and the Pfaffian line bundle over $C^{\infty}(\Sigma,M)$, namely that a choice of a string structure determines a trivialization of the Pfaffian line bundle, has been conjectured by Stolz and Teichner, and proved by Bunke \cite{bunke1}  using the notion of geometric string  structures of \cite{waldorf8}.

\section{Topological T-duality}

\label{tduality}

Particularly interesting target spaces for string theory are the total spaces of principal torus bundles. String theories on these target spaces can be equivalent to string theories on different principal torus bundles, in a way that metrics, B-fields, and dilaton fields are mixed up \cite{Buscher1987}. Such an equivalence is called \emph{T-duality}. The fully-fledged exact mathematical formulation of T-duality, including all topological and differential-geometric  information, is not yet known.

\emph{Topological} T-duality has been invented to study the underlying topological aspects alone. In this context, T-duality can be defined as follows \cite{Bunke2006a}. Let $E$ and $\hat E$ be principal $\mathbb{T}^{n}$-bundles over $M$, where $\mathbb{T}^{n}=U(1) \times 
\ldots \times U(1)$, and let $\mathcal{G}$ and $\widehat{\mathcal{G}}$ be $U(1)$-bundle gerbes over $E$ and $\hat E$, respectively. The pairs $(E,\mathcal{G})$ and $(\hat E,\widehat{\mathcal{G}})$ are called \emph{topological T-backgrounds}, and the fibre product
\begin{equation*}
\xymatrix{& E \times_M \hat E \ar[dl] \ar[dr] \\ E \ar[dr] && \hat E \ar[dl] \\ & M}
\end{equation*}
is called the \emph{correspondence space}.
The two topological T-backgrounds are called \emph{T-dual}, if the pullbacks of the two bundle gerbes to the correspondence space are isomorphic, and an isomorphism $\mathcal{D}:p^{*}\mathcal{G} \to \hat p^{*}\widehat{\mathcal{G}}$ exists that has the so-called Poincar\'e property \cite{Bunke2006a}. 
The original motivation for this definition was the existence of a so-called Fourier-Mukai transformation, which yields an isomorphism between the twisted K-theories of both pairs.

One of the basic questions in this setting is to decide, if a given T-background has T-duals, and how the possibly many T-duals can be parameterized. To this end, we consider the Serre spectral sequence associated to the torus bundle, which comes with a filtration 
$\pi^{*}\mathrm{H}^3(M,\mathbb{Z}) 
= F_3\subset F_2 \subset F_1 \subset F_0=\mathrm{H}^3(E,\mathbb{Z})$. We classify T-backgrounds by the greatest $n$ such that the Dixmier-Douady class $[\mathcal{G}] \in \mathrm{H}^3(E,\mathbb{Z})$ is in $F_n$. A result of Bunke-Rumpf-Schick \cite{Bunke2006a} is that a T-background $(E,\mathcal{G})$ admits T-duals if and only if it is $F_2$. Further, up to isomorphism, possible choices are related by a certain action of the additive group $\mathfrak{so}(n,\mathbb{Z})$ of skew-symmetric  matrices $B\in \mathbb{Z}^{n\times n}$, where $n$ is the dimension of the torus.

The situation can be  reformulated and then improved using non-abelian gerbes \cite{Nikolause}. In this joint work with Thomas Nikolaus we have manufactured a Fr\'echet Lie 2-group $\TBB$ in such a way that the $\TBB$-bundle gerbes are precisely the $F_2$ T-backgrounds. As a crossed module, it is  $\Z^{n} \times C^{\infty}(\mathbb{T}^{n},U(1)) \to \R^{n}$, defined as $(m,\tau)\mapsto m$, and $a\in \R^{n}$ acts on $C^{\infty}(\T^{n},U(1))$ by translations modulo $\Z$. It forms a (non-central) extension 
\begin{equation*}
C^{\infty}(\T^{n},U(1)) \to \TBB \to \T^{n}_{dis}\text{.}
\end{equation*}
If $\mathcal{G}$ is a $\TBB$-bundle gerbe corresponding to a T-background $(E,\mathcal{G})$, then $\pi_0(\mathcal{G})$ is the underlying torus bundle $E$. 

Another, finite-dimensional Lie 2-group $\TD$ can be constructed as the central extension
\begin{equation*}
BU(1) \to \TD \to \T_{dis}^{2n}
\end{equation*}
that is classified by the class 
\begin{equation*}
\rho := \sum_{i=1}^{n} \pr_i^{*}c \cup \pr_{n+i}^{*}c  \in\mathrm{H}^4(\T^{2n},\Z) \text{,}
\end{equation*}
where $c\in \mathrm{H}^2(BU(1),\Z)$ is the universal first Chern class. The associated $\TD$-bundle gerbes are precisely all T-duality correspondences \cite{Nikolause}. Lie 2-group homomorphisms $L,R:\TD \to \TBB$ represent the projection to the left and the right ,,leg`` of the correspondence. The main advantage of this reformulation is that the $\mathfrak{so}(n,\Z)$-action can be implemented as a strict and fully coherent action on $\TD$. This way, our understanding of topological T-duality is formulated completely and coherently in the language of non-abelian bundle gerbes.

If a T-background is only $F_1$, then it does not have any T-duals; these are then called 
,,mysteriously missing`` \cite{Mathai2006} or ,,non-geometric`` T-duals \cite{Hull2007}. 
An approach via non-commutative geometry allows to define them as bundles of non-commutative tori \cite{Mathai2005a,Mathai2006a,Mathai2006}.
The fact that non-abelian bundle gerbes form a 2-stack provides an alternative \cite{Nikolause}. Indeed,  every $F_1$ background is locally $F_2$, and so has locally defined T-duals, related to the given $F_1$ T-background by locally defined T-duality correspondences. Over overlaps, these correspondences are related by the $\mathfrak{so}(n,\Z)$-action of Bunke-Rumpf-Schick. Since this action is fully coherent under our reformulation by non-abelian bundle gerbes, one can define a semi-direct product Lie 2-group
\begin{equation*}
\TFgeo := \TD \ltimes \mathfrak{so}(n,\Z)\text{.}
\end{equation*}
Within the 2-stack of $\TFgeo$-bundle gerbes one can now \emph{glue} the locally defined T-duality correspondences along their $\mathfrak{so}(n,\Z)$-transformations on overlaps. This way, a globally defined $\TFgeo$-bundle gerbe is obtained, representing a totally new object called a \emph{half-geometric T-duality correspondence}. These new objects should be seen and studied as generalized target spaces for string theory, and may be seen as a realization of Hull's T-folds \cite{Hull2005}.

Another action of $\mathfrak{so}(n,\Z)$ on the Lie 2-group $\TBB$ can be defined, leading to another Lie 2-group
\begin{equation*}
\TB := \TBB \ltimes \mathfrak{so}(n,\Z)\text{.}
\end{equation*}  
The non-abelian $\TB$-bundle gerbes correspond precisely to the $F_1$ T-backgrounds \cite{Nikolause}. The left leg projection $L$ is $\mathfrak{so}(n,\Z)$-equivariant and hence induces a well-defined 2-group homomorphism $L: \TFgeo \to \TB$. In other words, half-geometric T-duality correspondences still have a well-defined ,,geometric`` left leg, but opposed to the theory of \cite{Bunke2006a} this left leg is now in the bigger class of $F_1$ T-backgrounds. It is shown in \cite{Nikolause} that $L: \TFgeo \to \TB$ induces a bijection on isomorphism classes of non-abelian gerbes. Thus, every $F_1$ T-background is the left leg of a uniquely defined half-geometric T-duality correspondence. They can hence be seen as the non-geometric T-duals of $F_1$ T-backgrounds. Thus, the higher geometry of non-abelian gerbes provides an alternative to non-commutative geometry.


\begin{footnotesize}

\bibliographystyle{sfb676}
\bibliography{sfb676_A4}

\providecommand{\href}[2]{#2}\begingroup\raggedright\begin{thebibliography}{10}

\bibitem{fuchs8}
J.~Fuchs, T.~Nikolaus, C.~Schweigert and K.~Waldorf, \emph{Bundle gerbes and
  surface holonomy},  in \emph{Proceedings of the 5th European Congress of
  Mathematics} (A.~Ran, H.~te~Riele and J.~Wiegerinck, eds.), pp.~167--197.
\newblock EMS, 2008.

\bibitem{Nikolaus:2010vq}
T.~Nikolaus and C.~Schweigert, \emph{{Equivariance In Higher Geometry}},
  \href{https://doi.org/10.1016/j.aim.2010.10.016}{\emph{Adv. Math.} {\bfseries
  226} (2011) 3367--3408}, [\href{https://arxiv.org/abs/1004.4558}{{\ttfamily
  1004.4558}}].

\bibitem{murray}
M.~K. Murray, \emph{Bundle gerbes}, {\emph{J. Lond. Math. Soc.} {\bfseries 54}
  (1996) 403--416}, [\href{https://arxiv.org/abs/dg-ga/9407015}{{\ttfamily
  dg-ga/9407015}}].

\bibitem{murray2}
M.~K. Murray and D.~Stevenson, \emph{Bundle gerbes: stable isomorphism and
  local theory}, {\emph{J. Lond. Math. Soc.} {\bfseries 62} (2000) 925--937},
  [\href{https://arxiv.org/abs/math/9908135}{{\ttfamily math/9908135}}].

\bibitem{stevenson1}
D.~Stevenson, \emph{The geometry of bundle gerbes}, Ph.D. thesis, University of
  Adelaide, 2000.
\newblock \href{https://arxiv.org/abs/math.DG/0004117}{{\ttfamily
  math.DG/0004117}}.

\bibitem{meinrenken1}
E.~Meinrenken, \emph{The basic gerbe over a compact simple {L}ie group},
  {\emph{Enseign. Math., II. S\'er.} {\bfseries 49} (2002) 307--333},
  [\href{https://arxiv.org/abs/math/0209194}{{\ttfamily math/0209194}}].

\bibitem{Waldorf:2007}
K.~Waldorf, \emph{{More morphisms between bundle gerbes}}, {\emph{Th.\ Appl.\
  Cat.} {\bfseries 18} (2007) 240--273},
  [\href{https://arxiv.org/abs/0702652}{{\ttfamily 0702652}}].

\bibitem{Baez2011}
J.~C. Baez and J.~Huerta, \emph{An invitation to higher gauge theory},
  {\emph{General Relativity and Gravitation} {\bfseries 43} (2011) 2335--2392},
  [\href{https://arxiv.org/abs/1003.4485}{{\ttfamily 1003.4485}}].

\bibitem{Saemann2014}
C.~Saemann and M.~Wolf, \emph{Non-abelian tensor multiplet equations from
  twistor space}, {\emph{Commun. Math. Phys.} {\bfseries 328} (2014) 527--544},
  [\href{https://arxiv.org/abs/1205.3108}{{\ttfamily 1205.3108}}].

\bibitem{Fiorenza2014}
D.~Fiorenza, H.~Sati and U.~Schreiber, \emph{Multiple {M}5-branes, string
  2-connections, and 7d non-abelian {C}hern-{S}imons theory}, {\emph{Adv.
  Theor. Math. Phys.} {\bfseries 18} (2014) 229--321},
  [\href{https://arxiv.org/abs/1201.5277}{{\ttfamily 1201.5277}}].

\bibitem{Jurco2015}
B.~Jurco, C.~Saemann and M.~Wolf, \emph{Semistrict higher gauge theory},
  {\emph{J. High Energy Phys.} {\bfseries 04} (2015) 087},
  [\href{https://arxiv.org/abs/1403.7185}{{\ttfamily 1403.7185}}].

\bibitem{Gukovb}
S.~Gukov and A.~Kapustin, \emph{Topological quantum field theory, nonlocal
  operators, and gapped phases of gauge theories},
  \href{https://arxiv.org/abs/1307.4793}{{\ttfamily 1307.4793}}. Preprint.

\bibitem{Freed2016a}
D.~S. Freed, M.~J. Hopkins and C.~Teleman, \emph{Loop groups and twisted
  {K}-theory {I}}, {\emph{J. Topology} {\bfseries 4} (2016) 737--798},
  [\href{https://arxiv.org/abs/0711.1906}{{\ttfamily 0711.1906}}].

\bibitem{Nikolause}
T.~Nikolaus and K.~Waldorf, \emph{Higher geometry for non-geometric {T}-duals},
   \href{https://arxiv.org/abs/1804.00677}{{\ttfamily 1804.00677}}. Preprint.

\bibitem{aschieri}
P.~Aschieri, L.~Cantini and B.~Jurco, \emph{Nonabelian bundle gerbes, their
  differential geometry and gauge theory}, {\emph{Commun. Math. Phys.}
  {\bfseries 254} (2005) 367--400},
  [\href{https://arxiv.org/abs/hep-th/0312154}{{\ttfamily hep-th/0312154}}].

\bibitem{breen1}
L.~Breen and W.~Messing, \emph{Differential geometry of gerbes}, {\emph{Adv.
  Math.} {\bfseries 198} (2005) 732--846},
  [\href{https://arxiv.org/abs/math.AG/0106083}{{\ttfamily math.AG/0106083}}].

\bibitem{Nikolaus}
T.~Nikolaus and K.~Waldorf, \emph{Four equivalent versions of non-abelian
  gerbes}, \href{https://doi.org/DOI 10.2140/pjm.2013.264.355}{\emph{Pacific J.
  Math.} {\bfseries 264} (2013) 355--420},
  [\href{https://arxiv.org/abs/1103.4815}{{\ttfamily 1103.4815}}].

\bibitem{pries2}
C.~Schommer-Pries, \emph{Central extensions of smooth 2-groups and a
  finite-dimensional string 2-group}, {\emph{Geom. Topol.} {\bfseries 15}
  (2011) 609--676}, [\href{https://arxiv.org/abs/0911.2483}{{\ttfamily
  0911.2483}}].

\bibitem{Nikolausa}
T.~Nikolaus and K.~Waldorf, \emph{Lifting problems and transgression for
  non-abelian gerbes}, {\emph{Adv. Math.} {\bfseries 242} (2013) 50--79},
  [\href{https://arxiv.org/abs/1112.4702}{{\ttfamily 1112.4702}}].

\bibitem{schreiber3}
U.~Schreiber and K.~Waldorf, \emph{Parallel transport and functors}, {\emph{J.
  Homotopy Relat. Struct.} {\bfseries 4} (2009) 187--244},
  [\href{https://arxiv.org/abs/0705.0452v2}{{\ttfamily 0705.0452v2}}].

\bibitem{schreiber5}
U.~Schreiber and K.~Waldorf, \emph{Smooth functors vs. differential forms},
  {\emph{Homology, Homotopy Appl.} {\bfseries 13} (2011) 143--203},
  [\href{https://arxiv.org/abs/0802.0663}{{\ttfamily 0802.0663}}].

\bibitem{schreiber6}
U.~Schreiber and K.~Waldorf, \emph{Local theory for 2-functors on path
  2-groupoids}, \href{https://doi.org/10.1007/s40062-016-0140-4}{\emph{J.
  Homotopy Relat. Struct.} (2016) 1--42},
  [\href{https://arxiv.org/abs/1303.4663}{{\ttfamily 1303.4663}}].

\bibitem{schreiber2}
U.~Schreiber and K.~Waldorf, \emph{Connections on non-abelian gerbes and their
  holonomy}, {\emph{Theory Appl. Categ.} {\bfseries 28} (2013) 476--540},
  [\href{https://arxiv.org/abs/0808.1923}{{\ttfamily 0808.1923}}].

\bibitem{bartels}
T.~Bartels, \emph{2-bundles and higher gauge theory}, Ph.D. thesis, University
  of California, Riverside, 2004.
\newblock \href{https://arxiv.org/abs/math/0410328}{{\ttfamily math/0410328}}.

\bibitem{Waldorf2016}
K.~Waldorf, \emph{A global perspective to connections on principal 2-bundles},
  \href{https://arxiv.org/abs/1608.00401}{{\ttfamily 1608.00401}}. to appear in
  Forum Math.

\bibitem{Waldorf2017}
K.~Waldorf, \emph{Parallel transport in principal 2-bundles},
  \href{https://arxiv.org/abs/1704.08542}{{\ttfamily 1704.08542}}. Preprint.

\bibitem{lerman1}
E.~Lerman, \emph{Orbifolds as stacks?}, {\emph{Enseign. Math} {\bfseries 56}
  (2010) 315--363}, [\href{https://arxiv.org/abs/0806.4160}{{\ttfamily
  0806.4160}}].

\bibitem{Moerdijk1997}
I.~Moerdijk and D.~A. Pronk, \emph{Orbifolds, sheaves and groupoids},
  {\emph{K-Theory} {\bfseries 12} (1997) 3--21}.

\bibitem{Moermrc}
I.~Moerdijk and J.~Mr{\v c}un, \emph{Introduction to foliations and Lie
  groupoids}.
\newblock Cambridge University Press, 2003.

\bibitem{nikolaus1}
T.~Nikolaus, \emph{{\"A}quivariante {G}erben und {A}bstieg}, . Diploma thesis,
  Universit{\"a}t Hamburg, 2009.

\bibitem{Gawedzki:2003pm}
K.~Gaw\c{e}dzki and N.~Reis, \emph{{Basic gerbe over nonsimply connected
  compact groups}},
  \href{https://doi.org/10.1016/j.geomphys.2003.11.004}{\emph{J. Geom. Phys.}
  {\bfseries 50} (2004) 28},
  [\href{https://arxiv.org/abs/math/0307010}{{\ttfamily math/0307010}}].

\bibitem{Krepski}
D.~Krepski, \emph{Basic equivariant gerbes on non-simply connected compact
  simple lie groups},  \href{https://arxiv.org/abs/1712.08294}{{\ttfamily
  1712.08294}}. Preprint.

\bibitem{Schreiber:2005mi}
U.~Schreiber, C.~Schweigert and K.~Waldorf, \emph{{Unoriented WZW models and
  holonomy of bundle gerbes}},
  \href{https://doi.org/10.1007/s00220-007-0271-x}{\emph{Commun. Math. Phys.}
  {\bfseries 274} (2007) 31--64},
  [\href{https://arxiv.org/abs/hep-th/0512283}{{\ttfamily hep-th/0512283}}].

\bibitem{gawedzki6}
K.~Gaw\c{e}dzki, R.~R. Suszek and K.~Waldorf, \emph{{WZW} orientifolds and
  finite group cohomology}, {\emph{Commun. Math. Phys.} {\bfseries 284} (2007)
  1--49}, [\href{https://arxiv.org/abs/hep-th/0701071}{{\ttfamily
  hep-th/0701071}}].

\bibitem{Gawedzki:2008um}
K.~Gaw\c{e}dzki, R.~R. Suszek and K.~Waldorf, \emph{{Bundle gerbes for
  orientifold sigma models}},
  \href{https://doi.org/10.4310/ATMP.2011.v15.n3.a1}{\emph{Adv. Theor. Math.
  Phys.} {\bfseries 15} (2011) 621--687},
  [\href{https://arxiv.org/abs/0809.5125}{{\ttfamily 0809.5125}}].

\bibitem{waldorf4}
K.~Waldorf, \emph{Algebraic structures for bundle gerbes and the
  {W}ess-{Z}umino term in conformal field theory}, Ph.D. thesis, Universit\"at
  Hamburg, 2007.

\bibitem{Hekmati}
P.~Hekmati, M.~K. Murray, R.~J. Szabo and R.~F. Vozzo, \emph{Real bundle
  gerbes, orientifolds and twisted {KR}-homology},
  \href{https://arxiv.org/abs/1608.06466}{{\ttfamily 1608.06466}}. Preprint.

\bibitem{Distler2011}
J.~Distler, D.~S. Freed and G.~W. Moore, \emph{Orientifold precis},  in
  \emph{Mathematical foundations of quantum field theory and perturbative
  string theory}, vol.~83, pp.~159--172.
\newblock AMS, 2011.
\newblock \href{https://arxiv.org/abs/0906.0795}{{\ttfamily 0906.0795}}.

\bibitem{freed5}
D.~S. Freed and G.~W. Moore, \emph{Setting the quantum integrand of
  {M}-theory}, {\emph{Commun. Math. Phys.} {\bfseries 263} (2006) 89--132},
  [\href{https://arxiv.org/abs/hep-th/0409135}{{\ttfamily hep-th/0409135}}].

\bibitem{Freed1986}
D.~S. Freed, \emph{Determinants, torsion, and strings}, {\emph{Commun. Math.
  Phys.} {\bfseries 107} (1986) 483--513}.

\bibitem{killingback1}
T.~Killingback, \emph{World sheet anomalies and loop geometry}, {\emph{Nuclear
  Phys. B} {\bfseries 288} (1987) 578}.

\bibitem{mclaughlin1}
D.~A. McLaughlin, \emph{Orientation and string structures on loop space},
  {\emph{Pacific J. Math.} {\bfseries 155} (1992) 143--156}.

\bibitem{stolz1}
S.~Stolz and P.~Teichner, \emph{What is an elliptic object?},  in
  \emph{Topology, geometry and quantum field theory}, vol.~308 of \emph{London
  Math. Soc. Lecture Note Ser.}, pp.~247--343.
\newblock Cambridge Univ. Press, 2004.

\bibitem{carey4}
A.~L. Carey, S.~Johnson, M.~K. Murray, D.~Stevenson and B.-L. Wang,
  \emph{Bundle gerbes for {C}hern-{S}imons and {W}ess-{Z}umino-{W}itten
  theories}, {\emph{Commun. Math. Phys.} {\bfseries 259} (2005) 577--613},
  [\href{https://arxiv.org/abs/math/0410013}{{\ttfamily math/0410013}}].

\bibitem{waldorf8}
K.~Waldorf, \emph{String connections and {C}hern-{S}imons theory},
  {\emph{Trans. Amer. Math. Soc.} {\bfseries 365} (2013) 4393--4432},
  [\href{https://arxiv.org/abs/0906.0117}{{\ttfamily 0906.0117}}].

\bibitem{Nikolausb}
T.~Nikolaus, C.~Sachse and C.~Wockel, \emph{A smooth model for the string
  group}, {\emph{Int. Math. Res. Not. IMRN} {\bfseries 16} (2013) 3678--3721},
  [\href{https://arxiv.org/abs/1104.4288}{{\ttfamily 1104.4288}}].

\bibitem{baez9}
J.~C. Baez, A.~S. Crans, D.~Stevenson and U.~Schreiber, \emph{From loop groups
  to 2-groups}, {\emph{Homology, Homotopy Appl.} {\bfseries 9} (2007)
  101--135}, [\href{https://arxiv.org/abs/math.QA/0504123}{{\ttfamily
  math.QA/0504123}}].

\bibitem{Waldorf}
K.~Waldorf, \emph{A construction of string 2-group models using a
  transgression-regression technique},  in \emph{Analysis, Geometry and Quantum
  Field Theory} (C.~L. Aldana, M.~Braverman, B.~Iochum and C.~Neira-Jim\'enez,
  eds.), vol.~584 of \emph{Contemp. Math.}, pp.~99--115.
\newblock AMS, 2012.
\newblock \href{https://arxiv.org/abs/1201.5052}{{\ttfamily 1201.5052}}.
\newblock \href{https://doi.org/http://dx.doi.org/10.1090/conm/584/11588}{DOI}.

\bibitem{waldorfb}
K.~Waldorf, \emph{String geometry vs. spin geometry on loop spaces},
  \href{https://doi.org/10.1016/j.geomphys.2015.07.003}{\emph{J. Geom. Phys.}
  {\bfseries 97} (2015) 190--226},
  [\href{https://arxiv.org/abs/1403.5656}{{\ttfamily 1403.5656}}].

\bibitem{bunke1}
U.~Bunke, \emph{String structures and trivialisations of a {P}faffian line
  bundle}, {\emph{Commun. Math. Phys.} {\bfseries 307} (2011) 675--712},
  [\href{https://arxiv.org/abs/0909.0846}{{\ttfamily 0909.0846}}].

\bibitem{Buscher1987}
T.~Buscher, \emph{A symmetry of the string background field equations},
  {\emph{Phys. Lett. B} {\bfseries 194} (1987) 59--62}.

\bibitem{Bunke2006a}
U.~Bunke, P.~Rumpf and T.~Schick, \emph{The topology of {T}-duality for
  $t^n$-bundles}, {\emph{Rev. Math. Phys.} {\bfseries 18} (2006) 1103--1154},
  [\href{https://arxiv.org/abs/math/0501487}{{\ttfamily math/0501487}}].

\bibitem{Mathai2006}
V.~Mathai and J.~Rosenberg, \emph{On mysteriously missing {T}-duals, {H}-flux
  and the {T}-duality group},  in \emph{Differential geometry and physics},
  vol.~10 of \emph{Nankai Tracts Math.}, pp.~350--358.
\newblock World Sci. Publ., 2006.
\newblock \href{https://arxiv.org/abs/hep-th/0409073}{{\ttfamily
  hep-th/0409073}}.

\bibitem{Hull2007}
C.~Hull, \emph{Global aspects of {T}-duality, gauged sigma models and
  {T}-folds}, {\emph{J. High Energy Phys.} (2007) 057},
  [\href{https://arxiv.org/abs/hep-th/0604178}{{\ttfamily hep-th/0604178}}].

\bibitem{Mathai2005a}
V.~Mathai and J.~Rosenberg, \emph{{T}-duality for torus bundles with {H}-fluxes
  via noncommutative topology}, {\emph{Commun. Math. Phys.} {\bfseries 253}
  (2005) 705--721}, [\href{https://arxiv.org/abs/hep-th/0401168}{{\ttfamily
  hep-th/0401168}}].

\bibitem{Mathai2006a}
V.~Mathai and J.~Rosenberg, \emph{{T}-duality for torus bundles with {H}-fluxes
  via noncommutative topology. {II}. the high-dimensional case and the
  {T}-duality group}, {\emph{Adv. Theor. Math. Phys.} {\bfseries 10} (2006)
  123--158}, [\href{https://arxiv.org/abs/hep-th/0508084}{{\ttfamily
  hep-th/0508084}}].

\bibitem{Hull2005}
C.~Hull, \emph{A geometry for non-geometric string backgrounds}, {\emph{J. High
  Energy Phys.} {\bfseries 10} (2005) 65},
  [\href{https://arxiv.org/abs/0406102}{{\ttfamily 0406102}}].

\end{thebibliography}\endgroup

\end{footnotesize}


\end{document}